\documentclass{article}
\usepackage{amsmath,amsthm,amssymb,tikz,url,listings,subfig}

\title{Initial Analysis of a Simple Numerical Model that Exhibits Antifragile Behavior}
\author{Bryan A. Knowles\footnote{bryan.knowles951@topper.wku.edu}}
\begin{document}
	\maketitle

	\begin{abstract}
		I present a simple numerical model based on iteratively updating subgroups
		of a population, individually modeled by nonnegative real numbers,
		by a constant decay factor; however, at each iteration, one group
		is selected to instead be updated by a constant growth factor. I discover
		a relationship between these variables and their respective probabilities
		for a given subgroup, summarized as the variable $c$. When $c>1$, the subgroup
		is found to tend towards behaviors reminiscent of antifragility; when at
		least one subgroup of the population has $c\ge1$, the population as a whole
		tends towards significantly higher probabilities of ``living forever,''
		although it may first suffer a drop in population size as less robust, fragile
		subgroups ``die off.''

		In concluding, I discuss the limitations and ethics of such a model, notably
		the implications of when an upper limit is placed on the growth constant,
		requiring a population to facilitate an increase in the decay factor to lessen
		the impact of periods of failure.
	\end{abstract}

	\section{Introduction}
		Antifragilty is a growing area of research in complex systems, and classical
		examples such as the hydra--who is strengthen by stress, up to a point,
		simply by growing two new heads each time an old one is cut off--pervade
		the literature. My aim in this work is to provide as simple a numerical
		model as possible that exhibits the nonlinear overcompensations expected
		of an antifragile system.

		The concept is straightforward: a population, say of rabbits, is composed
		of several subgroups, say of different colors. At each time step, such
		as generations, the population as a whole is exposed to a stressor from
		a set of possible stressors, such as a potential predator. Each subgroup
		will decrease in number except for at most one--this subgroup may be well
		adapted to this stressor and actually benefit from its presence, increasing
		in number.

		As time goes on, the population size as a whole will decrease until, if any
		such subgroups exist, the growth of the more often well adapted subgroups will
		offset the loss of the less adapted subgroups; eventually, it is hoped, the
		growth of these well adapted groups will overcompensate for the others,
		leading the population size to again grow as a whole. This is not unlike
		the self-healing material considered in \cite{jones14}, specifically those
		that ``borrow from areas of less stress to fortify areas under more stress.''

	\section{Model}
		I define a population $P$ as a set of positive real numbers $H$. By
		$H_i$ I denote the $i$th subgroup of the population, and by the population
		size I refer to the sum of the elements of $H$. I define a stressor
		$S$ as a set of random variables such that $S_t$ is a random group label $i$ at
		discrete time step $t$. Relatedly, I define a set of random variables $X$ such
		that $X_{it} = \beta$ iff $S_t=i$; otherwise, $X_{it} = \alpha$. By $\alpha$
		and $\beta$ I denote, respectively, a decay factor constant in $(0, 1)$
		and a growth factor constant in $(1, \infty)$.

		The model proceeds as follows: at each time step, one subgroup is selected
		of the population at random (via $S_t$); the selected subgroup grows by the factor $\beta$,
		and the other groups each decay by the factor $\alpha$; if a subgroup ever
		goes below $1.0$, it is considered to have ``died off'' and it is set to $0.0$;
		if the population size goes below $1.0$, then all subgroups must have died
		off, and the model halts; otherwise, it continues until a predetermined number
		of time steps.

		Note that no assumptions are made about the probability distributions
		of $S_t$ and, subsequently, $X_{it}$, other than that all $S_t$ are assumed
		to follow the same distribution. This means that some subgroups may be
		selected more often than others.

		With this model, I am interested in the longterm behavior of the population
		size with respects to a given stressor: iff the population is fragile to that
		stressor, then it will tend to die off; iff robust, it will tend to an equilibrium;
		and iff {\em antifragile}, it will tend to grow infinitely, where an antifragile
		system is loosely defined as ``a system that becomes stronger when stressed''
		\cite{jones14}, and in this particular application, a group exhibiting a nonlinear
		increase in group size over time in response to a stressor, i.e.,
		set of random group labels.

	\section{Analysis}
		Let $H_i(t)$ denote $H_i$ at time step $t$, that is, after $t$ iterations of
		the model; similarly, let $P(t)$ denote the population size at time step $t$.
		Then, by $(1)$-$(4)$, the expected value of $H_i(t)$ is found
		to be an exponential growth or decay function, dependant on whether the geometric
		mean of $X_i$ is, respectively, in $(1, \infty)$ or $(0, 1)$. The expected
		value of $P(t)$, by $(5)$, is simply a sum of such functions.

		\begin{align}
			H_i(t) &= \alpha^n \beta^m H_i(0), n+m = t & \text{Definition} \\
			       &= (\alpha^{n/t} \beta^{m/t})^t H_i(0) \\
			\lim_{t\to\infty} H_i(t) &= (\alpha^{Pr[X_i = \alpha]} \beta^{Pr[X_i = \beta]})^t H_i(0) & \text{Law of Large Numbers} \\
			E[H_i(t)] &= G[X_i]^t H_i(0), G[\cdot] = \text{geo. mean} & \text{Expected Value} \\
			E[P(t)]   &= \Sigma_i E[H_i(t)]
		\end{align}

		Let $\beta$ be chosen to be a function of $\alpha$, the distribution of $X_i$, and
		a constant $c$ such that $\beta = \alpha^{-c Pr[X_i=\alpha]/Pr[X_i=\beta]}$
		and $c > 0$. It is found by $(6)$-$(9)$ that the fragility of a subgroup can be
		determined solely by its value for $c$ in this function for $\beta$.

		\begin{align}
			H_i(t) &= (\alpha^{n/t} \alpha^{-cn/t})^t H_i(0) = \alpha^{n-cn} H_i(0) & \text{Substitution from (2)} \\
			c < 1 &\implies H_i(t) < H_i(0) \implies G[X_i] < 1 & \text{Fragile} \\
			c = 1 &\implies H_i(t) = H_i(0) \implies G[X_i] = 1 & \text{Robust} \\
			c > 1 &\implies H_i(t) > H_i(0) \implies G[X_i] > 1 & \text{Antifragile}
		\end{align}

		This implies that $E[H_i(t)]$ for each fragile subgroup $i$ is a decay
		function and will tend towards zero, eventually dying off as it crosses
		below $1.0$; and that $E[H_i(t)]$ for each antifragile subgroup $i$ is a
		growth function and will tend towards $\infty$. By this approach though, $E[H_i(t)]$
		for each robust subgroup $i$ is a straight line, implying that both robust
		and antifragile configurations will ``live forever.''
		However, due to the model's high variance, any subgroup can experience
		a long sequence of $\alpha$s, the population dying off as a whole out
		of ``bad luck.''

		So, let $L[\cdot]$ denote the lifespan of a variable, equivalently, the expected
		number of time steps before the variable first falls below $1.0$. Next,
		for a given subgroup $i$, consider the sequence of $\alpha$s and $\beta$s
		in $X_{it}$, ordered by $t$. If we define $X_{it}$ for time steps only during
		which $i$ has not yet died off, then obviously not all possible orderings
		of $\alpha$s and $\beta$s are possible; for example, a sequence of $n$ $\alpha$s
		followed by a single $\beta$ is not possible when $\alpha^n H_i(0) < 1.0$
		because the subgroup would have already died off before reaching the $\beta$.

		Let $K$ be the random variable of the number of $\beta$s in $X_{it}$
		for a given subgroup $i$. Let the initial health constant
		$w=\log_\alpha(H_i^{-1})+1$ represent the initial number of $\alpha$s needed
		for subgroup $i$ to die off and let the compensatory health constant
		$w'=\log_\alpha(\beta^{-1})$ represent the number of extra $\alpha$s needed
		for that subgroup to die off after that subgroup ``sees'' one $\beta$.

		Therefore, $L[H_i] = w+E[K]w'+E[K]$, that is, the sum of the initial
		health, compensatory health for each $\beta$ seen, and the number of $\beta$s
		seen themselves. By $(10)$-$(15)$, the probability mass function
		of $K$ is found in a manner similar to that of the binomial distribution.

		\begin{align}
			L[H_i] &\in \{w + kw' | k=0,1,...\} \\
			\text{Let } K &= (L[H_i] - w) / w' = k \\
			      Pr[K=0] &= Pr[X_i = \alpha]^w \\
			      Pr[K=k] &= Pr[X_i = \alpha]^{w+kw'} Pr[X_i = \beta]^{k} C(k) \\
			\text{Where } C(k) &= \text{valid orderings of $w+kw'$ $\alpha$s and $k$ $\beta$s} \\
			                     &= {\textstyle \binom{w+(k-1)w'+(k-1)}{k} - \Sigma_{i=0}^{k-2} C(i)
			                      \binom{(k-1-i)w'+(k-1-i)}{k-i}}
		\end{align}

		Figure 1 examines the commulative distribution function (CDF) of $K$ under several different
		configurations, illustrating that the behavior of the distribution
		is heavily influenced by $c$: when $c>1$, that is the subgroup is antifragile,
		the CDF quickly converges to a value less than $1.0$; and when $c<1$, that is the
		subgroup is fragile, the CDF quickly converges to $1.0$. This general behavior is
		irrespective of $Pr[X_i=\alpha]$ and $w$.

		\begin{figure}
			\centering
			\subfloat{\includegraphics[width=0.50\textwidth]{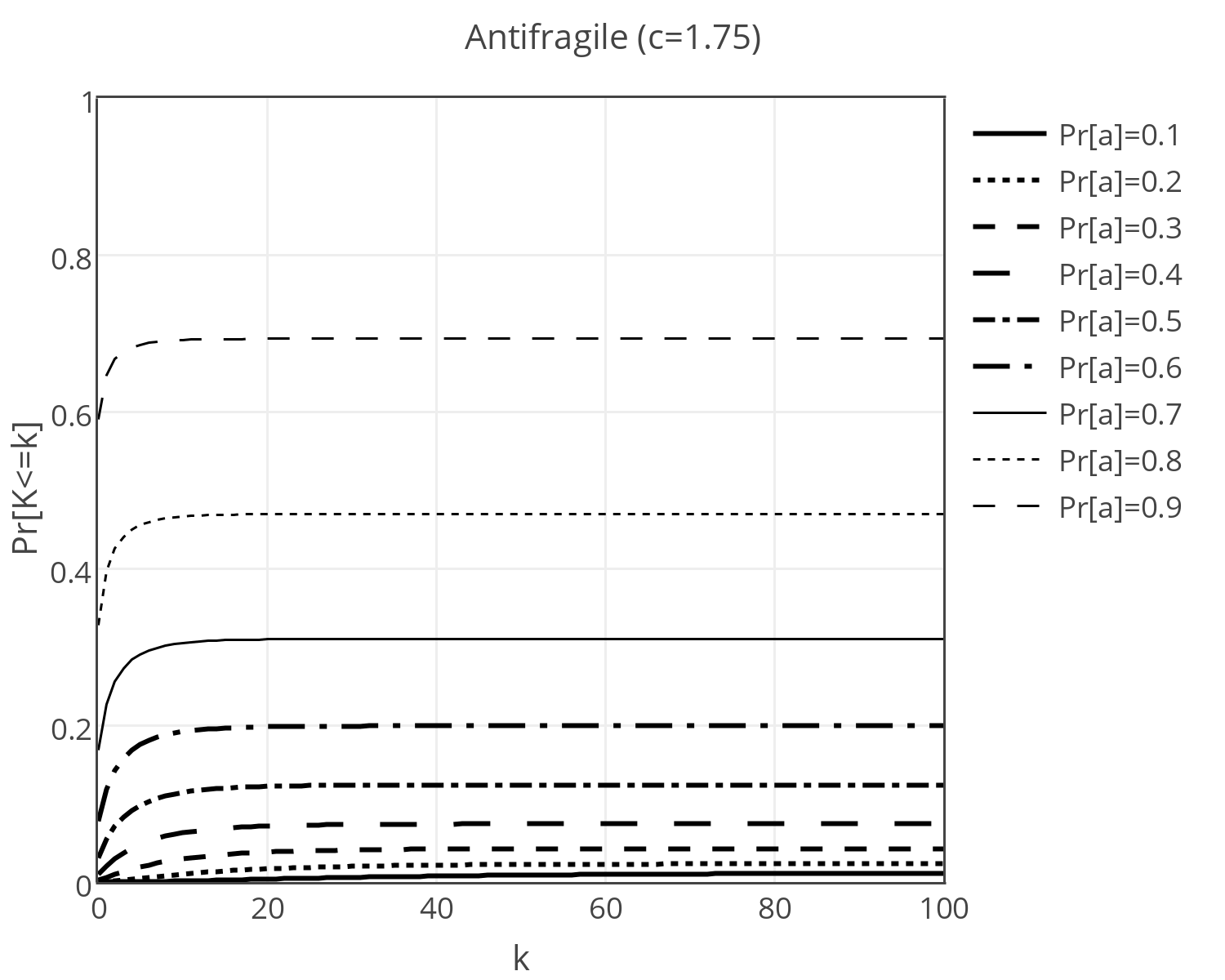}}
			\subfloat{\includegraphics[width=0.50\textwidth]{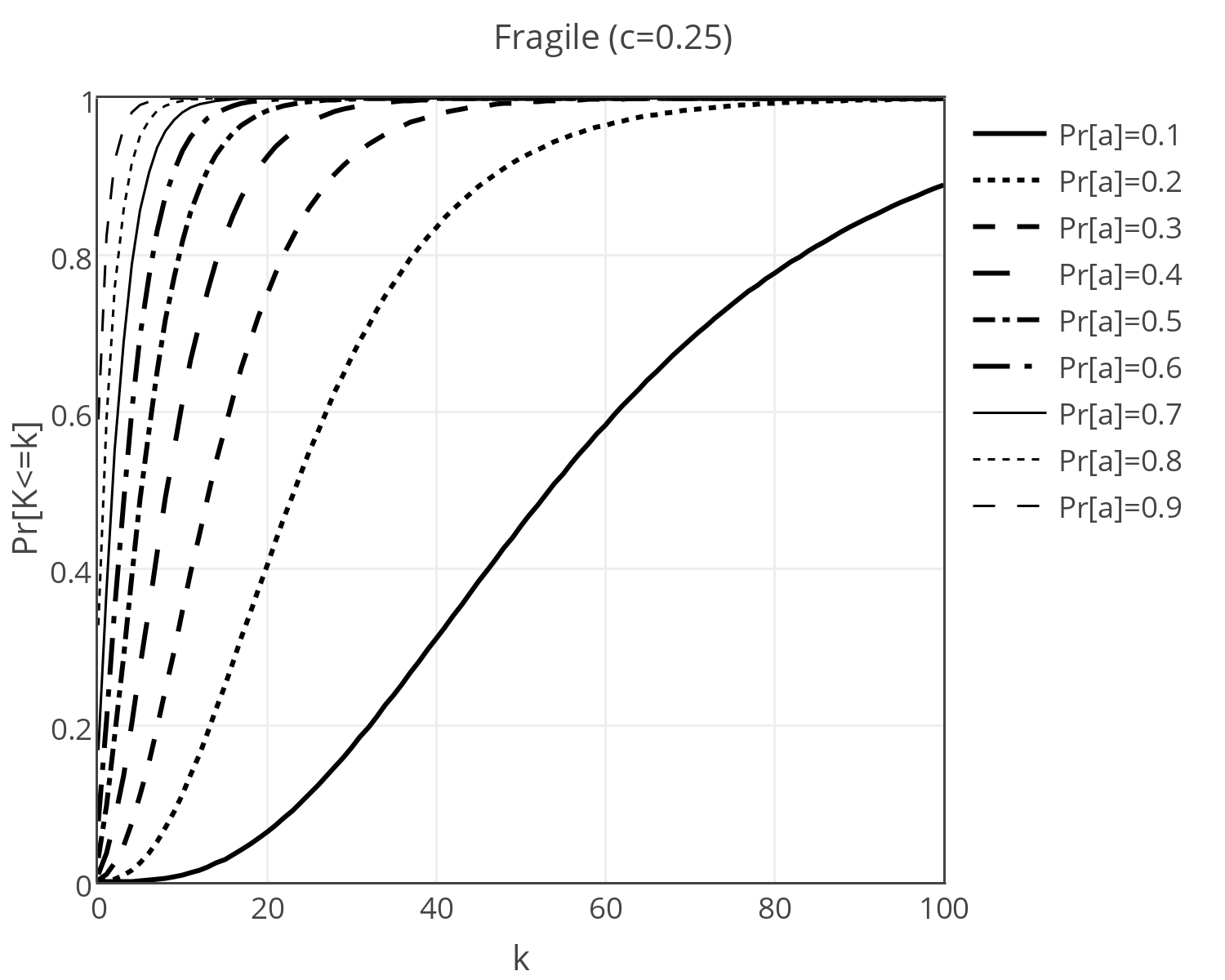}}\\
			\subfloat{\includegraphics[width=0.50\textwidth]{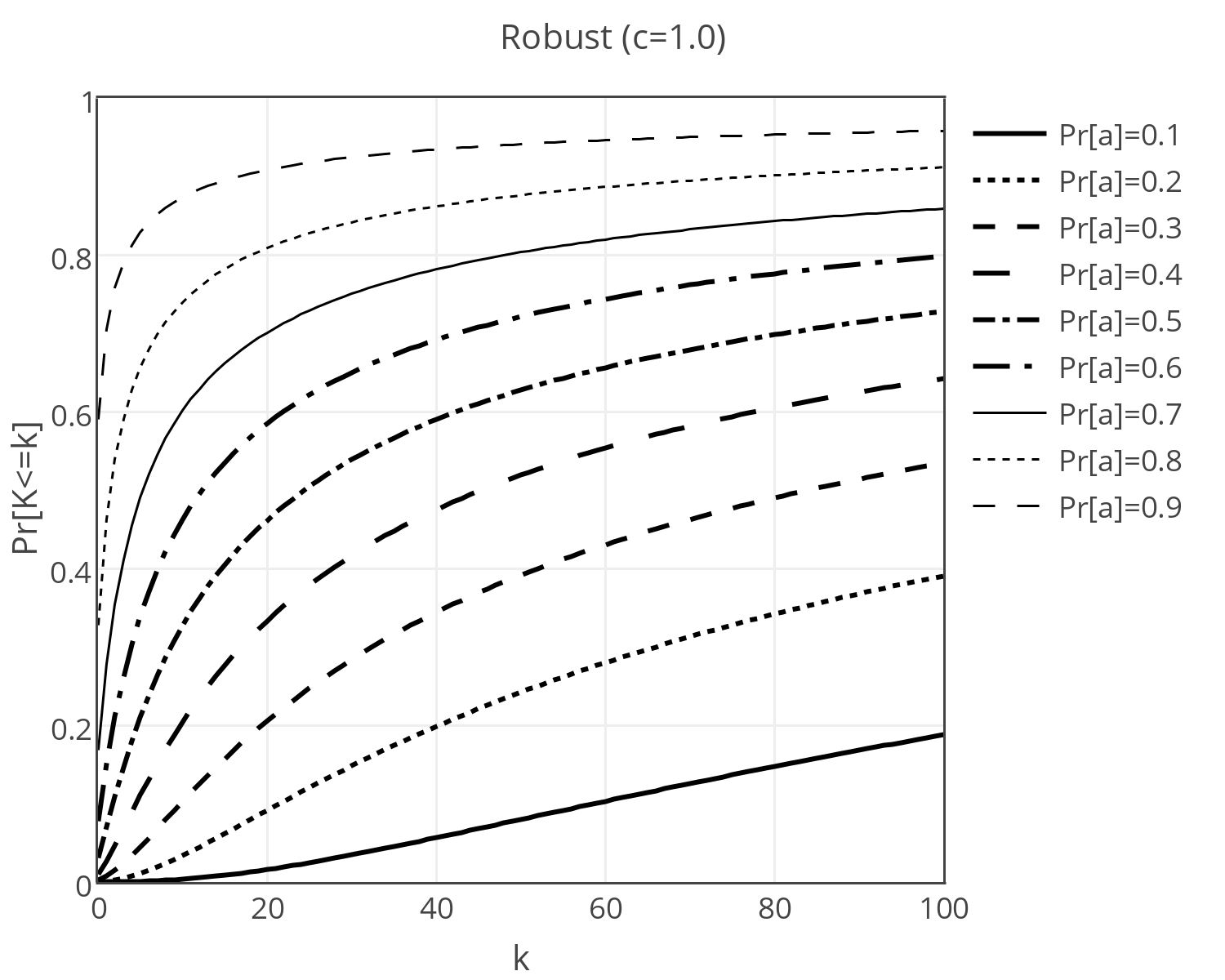}}
			\caption{CDFs of several $K$s: antifragile (top left),
			fragile (top right), and robust (bottom), with $Pr[X_i=\alpha]$ from $10\%$ to $90\%$,
			$k$ from $0$ to $100$, $w=5$, and $w'=1$ for each.}
			\label{fig:CDF}
		\end{figure}

		Because $Pr[K=k]$ represents the probability that subgroup $i$
		will see exactly $k$ $\beta$s and then die off and $Pr[K<\infty]$
		represents the probability that the subgroup will die off at all,
		the behavior of the CDF implies that an antifragile subgroup has a
		significant probability of ``living forever,'' whereas fragile subgroups
		have almost none and robust subgroups are not as straightforward to predict.

	\section{Verification}
		To verify the predictive power of the model, 5,000 simulations were run for each $k$ from
		0 to 100 where $\beta=2$, subgroups were labeled $1..5$, subgroup $i$ was selected
		for growth $\frac{i}{15}$ times, and $H_i(0)=10$ for each subgroup. The remaining
		parameters are found in $(16)$ - $(24)$ such that subgroup 3 is expected to be robust,
		subgroups 1 and 2 are expected to be fragile, and subgroups 4 and 5 are expected
		to be antifragile. Figure 2 illustrates the results, comparing calculated (predicted)
		CDFs for each subgroup and observed (simulated) survival rates. Predictions were
		made in a bignumber implementation of Julia 0.3.5 and simulations were run on
		a 64-bit floating point implementation in Python 2.7.8.

		\begin{align}
			\beta &= \alpha^{-Pr[X_i=\alpha]/Pr[X_i=\beta]} & \text{Robust} \\
			\beta &= 2, Pr[X_i=\beta] = i/15, i=3 & \text{Given} \\
			2 &= \alpha^{-4} \\
			  &\implies \alpha = 2^{-1/4} \approx 0.84089 \\
			w &= log_\alpha (H_i(0)^{-1}) + 1 & \text{Definition} \\
			H_i(0) &= 10 & \text{Given} \\
			w &= log_{2^{-1/4}}(10^{-1}) + 1 \approx 14.2877123 \\
			w' &= log_{\alpha}(\beta^{-1}) & \text{Definition} \\
			w' &= log_{2^{-1/4}}(2^{-1}) = 4
		\end{align}

		Figure 2 also illustrates the behavior of a population size over time for
		a model configuration with 100 subgroups, $H_i(0)=10$ for each subgroup,
		$\beta=1.5$, $\alpha=0.995$, and $t\le200$. I compared two methods for
		implementing $S_t$: an iterative method where
		$S_t=(t \mod 51)+(\left\lfloor\frac{t}{51}\right\rfloor \mod 51)$; and a
		random method where $S_t=\text{uniform}(0,50)+\text{uniform}(0,49)$. Note
		the clear global convexity of the population size over time--a requirement of
		antifragility--, although this convexity is composed of several local periods
		of concavity in the iterative example. In the random examples similar patterns
		can be noted, although they are not as pronounced.

		\begin{figure}
			\centering
			\subfloat{\includegraphics[width=0.50\textwidth]{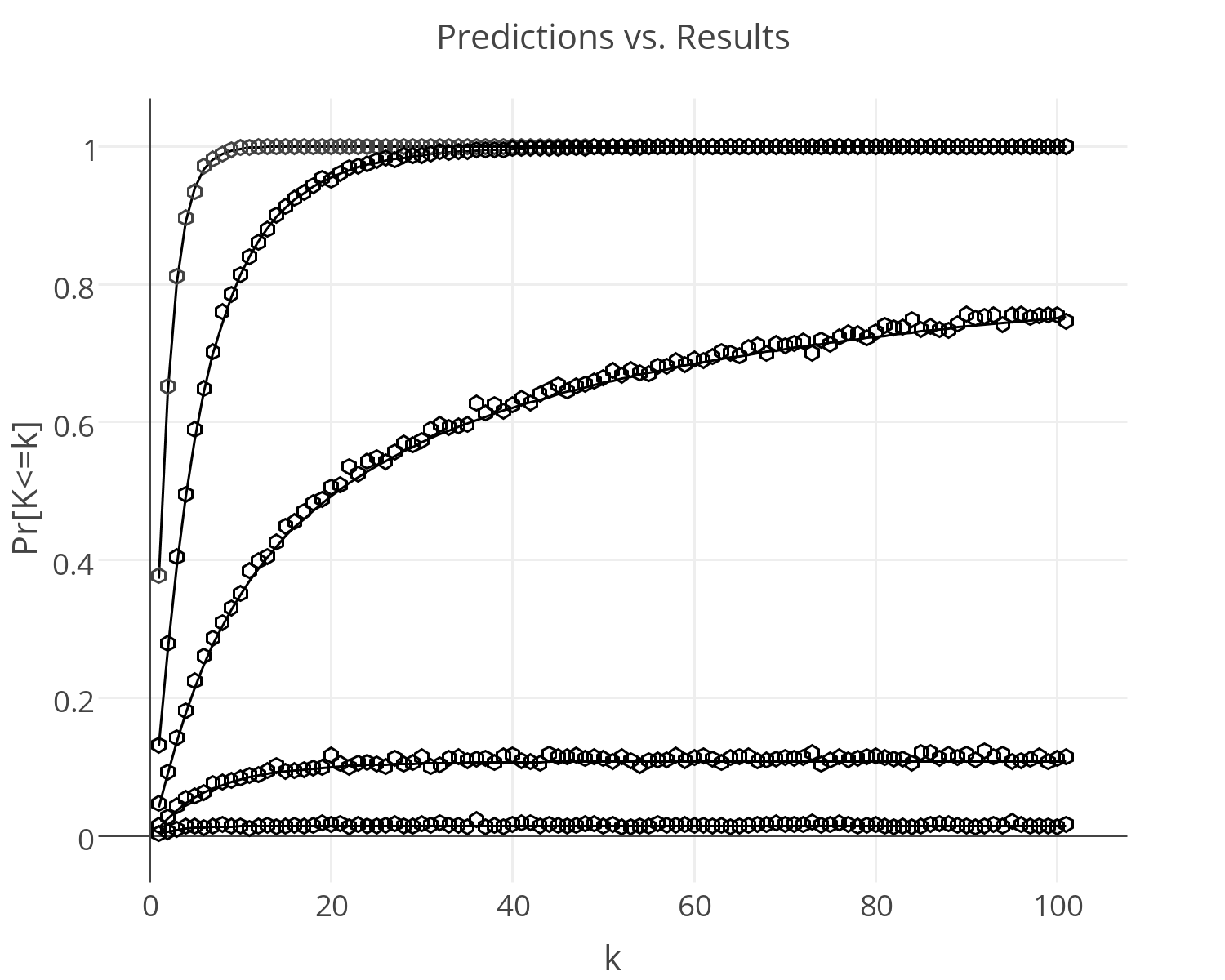}}\\
			\subfloat{\includegraphics[width=0.50\textwidth]{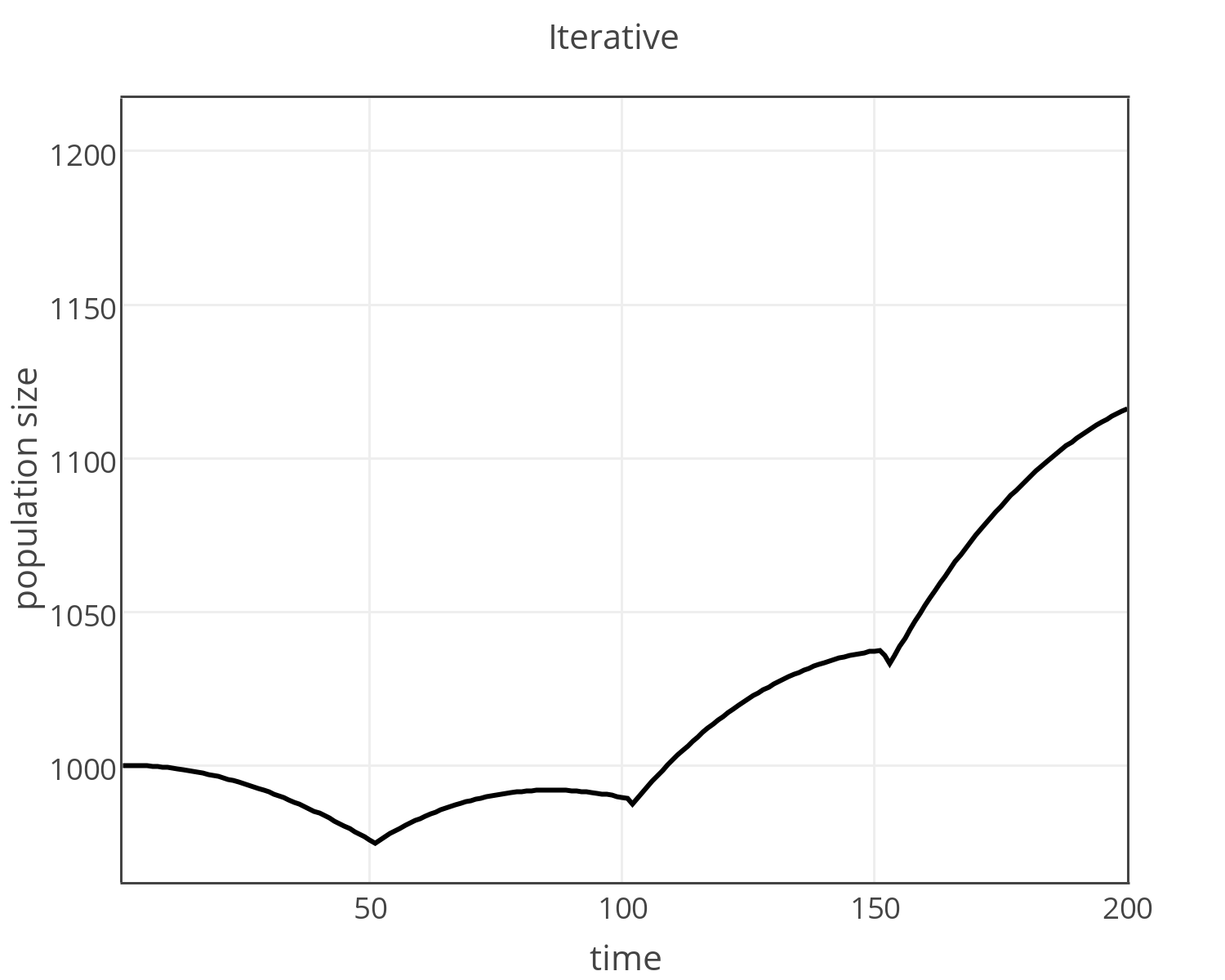}}
			\subfloat{\includegraphics[width=0.50\textwidth]{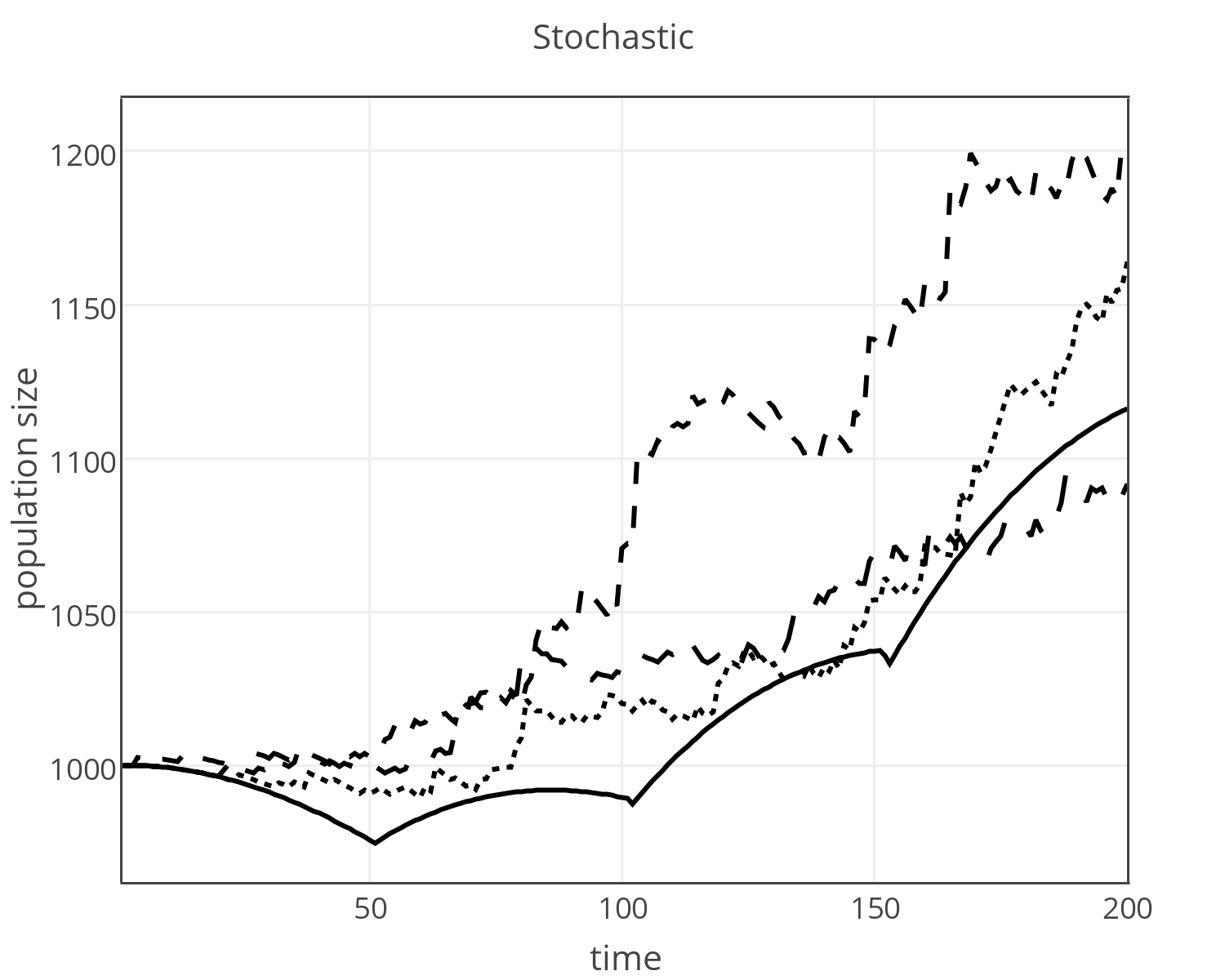}}
			\caption{Comparison (top) of predicted CDFs (lines) and observed
			survival rates (points) for subgroups 1 (top line) through 5 (bottom line).
			Also, population sizes over time using an iterative method to select
			subgroups (bottom left) and a random method (bottom right). The iterative data is
			shown on the right alongside three random runs' data for quick comparison.}
			\label{fig:Pt}
		\end{figure}

	\section{Conclusion}
		I have presented and analyzed a simple model that exhibits antifragility. In this
		model, the population as a whole is exposed to a stressor, represented by a set
		of random variables. At each time step, exactly one subgroup of the population
		benefits from this stressor, growing in size as the others shrink. In the long
		run behavior of the model, the weaker (fragile) subgroups of the population die
		off, represented as an exponential decay function while the stronger (robust or
		antifragile) subgroups either maintain an approximate equilibrium or grow infinitely
		in size.

		The model either contains no or masks with its simplicity any feedback
		loops, posited in \cite{bakhouyaa14} as necessary to produce a stable system:
		``positive feedback alone pushes the system beyond its limits and, eventually
		out of control, while negative feedback alone prevents the system from reaching
		its optimal behavior,'' a behavior echoed here in the tendancy of fragile groups
		to die off, antifragile groups to grow ``infinitely,'' and robust groups to
		rest somewhere in between. Without loops of any kind, then, the proposed model
		is instead little more than a set of functions with tendancies to grow or decay
		in response to a set of random variables.

		Yet does this model still suggest brute survival of the fittest? Yes, but only in cases
		where it is directly applicable, and these applications presuppose a static
		longterm stress environment and complete independence of population subgroups,
		neither of which are typical givens in complex systems such as social networks
		and biological ecosystems. This is, as it has been characterized here, the
		model does not support a stressor that \emph{changes suddenly}, presenting
		an opportunity for the model to suffer the ``turkey fallacy'' \cite{taleb14},
		where the system is fragile to abrupt environmental changes. Furthermore,
		because of the aforementioned limitations on the model, a general measure of
		antifragility cannot be given, albeit a general \emph{sense} of the term is.

		However, it does demonstrate a relationship between periods of failure, represented
		by the decay constant $\alpha$, and periods of overcompensation, representated by
		the growth constant $\beta$. If an upper limit is placed on the ability of a population
		to overcompensate from a stressor, as one would expect is often the
		case in natural systems, then once that limit has been reached the population's
		only choice to improve its robustness is to raise the decay factor--that is,
		increase $\alpha$ through some facility such that during periods of failure
		when overcompensation is not possible the impact of the stressor is not as
		severe.

	\bibliographystyle{plain}
	\bibliography{references.bib}

\end{document}